\documentclass[12pt]{article}
\usepackage{euscript}
\usepackage{amsmath}
\usepackage{amssymb}
\usepackage{cite}
\usepackage{color}

\def\mathswitchr#1{\relax\ifmmode{\mathrm{#1}}\else$\mathrm{#1}$\fi}

\newcommand {\pslash}{\hbox{$\not\hbox{\kern-2.3pt $p$}$}}

\def\rQCED{{\rm QCED}}

\def\alf1{ {\alpha\over\pi} }
\def\rQCD{{\rm QCD}}
\begin{document}
\begin{flushleft}
BU-HEPP-05-05\\July, 2005
\end{flushleft}
\vskip 2.0cm
\begin{center}
{\Large\bf QED$\otimes$QCD Exponentiation and Shower/ME Matching at the LHC$^{\dagger}$}\\
\vspace{10mm}
B.F.L. Ward\\

Department of Physics, Baylor University, Waco, TX, USA
\vskip .5cm


\vskip .5cm
S. A. Yost\\

Department of Physics, Baylor University, Waco, TX, USA \\
\vspace{5mm}
{\bf Abstract}\\
\end{center}
We present the elements of QED$\otimes$QCD
exponentiation and its interplay with shower/ME matching
in precision LHC physics scenarios. Applications to single heavy gauge
boson production at hadron colliders are illustrated.\\

\vspace{20mm}
\centerline{To appear in Proceedings of the HERA-LHC Workshop, 2005}
\vspace{10mm}
\renewcommand{\baselinestretch}{0.1}
\footnoterule
\noindent
{\footnotesize
\begin{itemize}
\item[${\dagger}$]
Work partly supported by US DOE grant DE-FG02-05ER41399 and
by NATO grant PST.CLG.980342.
\end{itemize}
}
\newpage
In the LHC environment, precision predictions for the effects 
of multiple gluon and multiple photon radiative processes will be needed to realize the true potential of the attendant physics program. For example,
while the current precision tag for the luminosity at FNAL is at the $\sim 7\%$
level~\cite{fnallum}, the high precision requirements for the LHC dictate 
an experimental precision tag for the luminosity at 
the $2\%$ level~\cite{lhclum}. This means that the theoretical precision tag
requirement for the corresponding luminosity processes, such as
single W,Z production with the subsequent decay into light lepton pairs,
must be at the 1\% level in order not to spoil the over-all precision
of the respective luminosity determinations at the LHC.
This theoretical precision tag means that multiple gluon and multiple
photon radiative effects in the latter processes must be controlled 
to the stated precision. With this objective in mind, we have developed
the theory of $QED\otimes QCD$ exponentiation to allow the simultaneous
resummation of the multiple gluon and multiple photon radiative
effects in LHC physics processes, to be realized ultimately by MC
methods on an event-by-event basis in the presence of parton showers
in a framework which allows us to systematically improve the accuracy
of the calculations without double-counting of effects in principle
to all orders in both $\alpha_s$ and $\alpha$.\par 

Specifically, the new $QED\otimes QCD$ exponentiation theory
is an extension of the $QCD$ exponentiation theory presented
in Refs.~\cite{qcdref}\footnote{We stress that the formal proof of 
exponentiation in non-Abelian gauge theories {\it in the eikonal approximation}
is given in Ref.~\cite{gatherall}. The results in Ref.~\cite{qcdref}
are in contrast {\it exact} but have an exponent that only contains
the leading contribution of the exponent in Ref.~\cite{gatherall}.}. We 
recall that in the latter references
it has been established that the following result holds for
a process such as $q+\bar q'\rightarrow V+n(G)+X\rightarrow
\bar{\ell} \ell'+n(g)+X$:
{\small
\begin{equation}
\begin{split}
d\hat\sigma_{\rm exp}&= \sum_n d\hat\sigma^n 
         =e^{\rm SUM_{IR}(QCD)}\sum_{n=0}^\infty\int\prod_{j=1}^n{d^3
k_j\over k_j}\\&\quad\int{d^4y\over(2\pi)^4}e^{iy\cdot(P_1+P_2-Q_1-Q_2-\sum k_j)+
D_\rQCD}\\
&*\tilde{\bar\beta}_n(k_1,\ldots,k_n){d^3P_2\over P_2^{\,0}}{d^3Q_2\over
Q_2^{\,0}}
\end{split}
\label{qcd}
\end{equation}}\noindent
where gluon residuals 
$\tilde{\bar\beta}_n(k_1,\ldots,k_n)$
, defined by Ref.~\cite{qcdref}, 
are free of all infrared divergences to all 
orders in $\alpha_s(Q)$. The functions
$SUM_{IR}(QCD), D_\rQCD$, together with 
the basic infrared functions 
$B^{nls}_{QCD},{\tilde B}^{nls}_{QCD},{\tilde S}^{nls}_{QCD}$ 
are specified in Ref.~\cite{qcdref}.
Here
$V=W^\pm,Z$,and $\ell = e,\mu,~\ell'=\nu_e,\nu_\mu ( e,\mu )$
respectively for $V=W^+ ( Z )$, and  
$\ell = \nu_e,\nu_\mu,~\ell'= e,\mu$ respectively for $V = W^-$.
We call attention to the essential
compensation between
the left over genuine non-Abelian IR virtual and real singularities
between $\int dPh\bar\beta_n$ and $\int dPh\bar\beta_{n+1}$ respectively
that really allows us to isolate $\tilde{\bar\beta}_j$ and distinguishes
QCD from QED, where no such compensation occurs.
The result in (\ref{qcd}) has been realized by Monte
Carlo methods~\cite{qcdref}.
See also Refs.~\cite{van1,van2,anas} for exact ${\cal O}(\alpha_s^2)$
and Refs.~\cite{baurall,ditt,russ} for exact ${\cal O}(\alpha)$
results on the W,Z production processes which we discuss here.
\par 

The new $QED\otimes QCD$ theory 
is obtained by simultaneously exponentiating the large 
IR terms in QCD and the exact IR divergent terms in QED, so that
we arrive at the new result
{\small
\begin{equation}
\begin{split}
d\hat\sigma_{\rm exp} &= e^{\rm SUM_{IR}(QCED)}\\
   &\sum_{{n,m}=0}^\infty\int\prod_{j_1=1}^n\frac{d^3k_{j_1}}{k_{j_1}} 
\prod_{j_2=1}^m\frac{d^3{k'}_{j_2}}{{k'}_{j_2}}
\int\frac{d^4y}{(2\pi)^4}\\&e^{iy\cdot(p_1+q_1-p_2-q_2-\sum k_{j_1}-\sum {k'}_{j_2})+
D_\rQCED} \\
&\tilde{\bar\beta}_{n,m}(k_1,\ldots,k_n;k'_1,\ldots,k'_m)\frac{d^3p_2}{p_2^{\,0}}\frac{d^3q_2}{q_2^{\,0}},
\end{split}
\label{qced}
\end{equation}}\noindent
where the new YFS~\cite{yfs,yfs1} residuals, defined in Ref.~\cite{CG1}, 
$\tilde{\bar\beta}_{n,m}(k_1,\ldots,k_n;k'_1,\ldots,k'_m)$, with $n$ hard gluons and $m$ hard photons,
represent the successive application
of the YFS expansion first for QCD and subsequently for QED. The
functions ${\rm SUM_{IR}(QCED)},D_\rQCED$ are determined
from their analogs ${\rm SUM_{IR}(QCD)},D_\rQCD$ via the
substitutions
{\small
\begin{eqnarray}
B^{nls}_{QCD} \rightarrow B^{nls}_{QCD}+B^{nls}_{QED}\equiv B^{nls}_{QCED},\cr
{\tilde B}^{nls}_{QCD}\rightarrow {\tilde B}^{nls}_{QCD}+{\tilde B}^{nls}_{QED}\equiv {\tilde B}^{nls}_{QCED}, \cr
{\tilde S}^{nls}_{QCD}\rightarrow {\tilde S}^{nls}_{QCD}+{\tilde S}^{nls}_{QED}\equiv {\tilde S}^{nls}_{QCED}
\label{irsub}
\end{eqnarray}}  
everywhere in expressions for the
latter functions given in Refs.~\cite{qcdref}.
The residuals $\tilde{\bar\beta}_{n,m}(k_1,\ldots,k_n;k'_1,\ldots,k'_m)$ 
are free of all infrared singularities
and the result in (\ref{qced}) is a representation that is exact
and that can therefore be used to make contact with parton shower 
MC's without double counting or the unnecessary averaging of effects
such as the gluon azimuthal angular distribution relative to its
parent's momentum direction.\par

In the respective infrared algebra (QCED) in (\ref{qced}), the average Bjorken $x$ values
\begin{eqnarray}
x_{avg}(QED)&\cong \gamma(QED)/(1+\gamma(QED))\nonumber\\
x_{avg}(QCD)&\cong \gamma(QCD)/(1+\gamma(QCD))
\nonumber
\end{eqnarray}
where
$\gamma(A)=\frac{2\alpha_{A}{\cal C}_A}{\pi}(L_s
-1)$, $A=QED,QCD$, with
${\cal C}_A=Q_f^2, C_F$, respectively, for 
$A=QED,QCD$ and the big log $L_s$, imply that
QCD dominant corrections happen an
order of magnitude earlier than those for QED. This means 
that the leading $\tilde{\bar\beta}_{0,0}$-level
gives already a good estimate of the size of the interplay between the
higher order QED and QCD effects which we will use to
illustrate (\ref{qced}) here.\par

More precisely, for the processes
$pp\rightarrow V +n(\gamma)+m(g)+X\rightarrow \bar{\ell} \ell'
+n'(\gamma)+m(g)+X$, where 
$V=W^\pm,Z$,and $\ell = e,\mu,~\ell'=\nu_e,\nu_\mu ( e,\mu )$
respectively for $V=W^+ ( Z )$, and  
$\ell = \nu_e,\nu_\mu,~\ell'= e,\mu$ respectively for $V = W^-$,
we have the usual formula
(we use the standard notation here~\cite{CG1}){\small
\begin{eqnarray}
d\sigma_{exp}(pp\rightarrow V+X\rightarrow \bar\ell \ell'+X') =\nonumber\\
\sum_{i,j}\int dx_idx_j F_i(x_i)F_j(x_j)d\hat\sigma_{exp}(x_ix_js),
\label{sigtot} 
\end{eqnarray}}\noindent
and we use the result in (\ref{qced}) here with semi-analytical
methods and structure functions from Ref.~\cite{mrst1}.
A Monte Carlo realization will appear elsewhere~\cite{elsewh}.
\par

We {\em do not} attempt in the {\it present} discussion
to replace HERWIG~\cite{herwig} 
and/or PYTHIA~\cite{pythia} --
we intend {\it here} 
to combine our exact YFS calculus with HERWIG and/or PYTHIA
{\em by using the latter to generate a parton shower starting from the initial $(x_1,x_2)$ point at factorization scale $\mu$ after this point is provided by the $\{F_i\}$}.
This combination of theoretical constructs can be 
systematically improved with
exact results order-by-order in $\alpha_s$, where  
currently the state of the art in such
a calculation is the work in Refs.~\cite{frixw}
which accomplishes the combination of an exact ${\cal O}(\alpha_s)$
correction with HERWIG. We note that, even in this latter result,
the gluon azimuthal angle is averaged in the combination. We note that
the recent alternative parton distribution function 
evolution MC algorithm in Refs.~\cite{jadskrz} 
can also be used in our theoretical
construction here. Due to its lack of the appropriate color coherence~\cite{mmm}, we do not consider ISAJET~\cite{isajet} here.\par

To illustrate how the combination with Pythia/Herwig can proceed,
we note that, for example, if we use a quark mass $m_q$ as our collinear limit
regulator, DGLAP~\cite{dglap} evolution of the structure functions allows us to
factorize all the terms that involve powers of the big log $L_c=\ln \mu^2/m_q^2-1$ in such a way that the evolved structure function contains the
effects of summing the leading big logs $L=\ln \mu^2/\mu_0^2$
where we have in mind that the evolution involves initial data at the
scale $\mu_0$. The result is therefore independent of $m_q$ for 
$m_q \downarrow 0$. In the context of the DGLAP theory, the factorization scale
$\mu$ represents the largest $p_\perp$ of the gluon emission included
in the structure function. In practice, when we use these structure functions
with an exact result for the residuals in (\ref{qced}), it means that
we must in the residuals omit the contributions from gluon radiation
at scales below $\mu$. This can be shown to amount in most cases to
replacing $L_s=\ln \hat{s}/m_q^2 -1 \rightarrow L_{nls}=\ln\hat{s}/\mu^2$
but in any case it is immediate how to limit the $p_T$ in the 
gluon emission~\footnote{ Here, we refer to both on-shell and off-shell
emitted gluons.} so that we do not double count effects.
In other words, we apply the standard QCD factorization of mass singularities
to the cross section in (\ref{qced}) in the standard way. We may do it
with either the mass regulator for the collinear singularities or with
dimensional regularization of such singularities -- the final result
should be independent of this regulator. This would in practice
mean the following: We first make an event with the formula
in (\ref{sigtot}) which would produce an initial beam state at $(x_1,x_2)$
for the two hard interacting partons at the factorization scale $\mu$
from the structure functions $\{F_j\}$ and a corresponding final state X
from the exponentiated cross section in $d\hat\sigma_{exp}(x_ix_js)$
; the standard Les Houches procedure~\cite{leshouches} of showering this event $(x_1,x_2,X)$ would then be used, employing backward evolution of the
initial partons. If we restrict the $p_T$ as we have
indicated above, there would be no double counting of effects.
Let us call this $p_T$ matching of the shower from the backward
evolution and the matrix elements in the QCED exponentiated cross section.\par

However, one could ask if it is possible to be more accurate
in the use of the exact result in (\ref{qced})? Indeed, it is.
Just as the residuals $\tilde{\bar\beta}_{n,m}(k_1,\ldots,k_n;k'_1,\ldots,k'_m)$are computed order by order in perturbation theory from the corresponding
exact perturbative results by expanding the exponents in (\ref{qced})
and comparing the appropriate corresponding coefficients of the
respective powers of $\alpha^n\alpha_s^m$, so too can the shower
formula which is used to generate the backward evolution be expanded
so that the product of the shower formula's perturbative expansion,
the perturbative expansion of the exponents in (\ref{qced}), and the
perturbative expansions of the residuals can be 
written as an over-all expansion in powers of $\alpha^n\alpha_s^m$
and required to match the respective calculated exact result
for given order. In this way, new shower subtracted
residuals, 
$\{\hat{\tilde{\bar\beta}}_{n,m}(k_1,\ldots,k_n;k'_1,\ldots,k'_m)\}$, 
are calculated that can be used for the entire gluon $p_T$
phase space with an accuracy of the cross section that
should in principle be improved compared with the first procedure for shower
matching presented above. Both approaches are under investigation.\par

Returning to the general discussion,
we compute, with and without QED, the ratio
$r_{exp}=\sigma_{exp}/\sigma_{Born}$,
where we do not use the narrow resonance approximation; for,
we wish to set a paradigm for precision heavy vector boson studies.
The formula which we use for $\sigma_{Born}$ is obtained from that in
(\ref{sigtot}) by substituting $d\hat\sigma_{Born}$ for $d\hat\sigma_{\rm exp}$
therein, where $d\hat\sigma_{Born}$ is the respective parton-level 
Born cross section. 
Specifically, we have from (\ref{qcd})
the $\tilde{\bar\beta}_{0,0}$-level result
\begin{equation}
\hat\sigma_{exp}(x_1x_2s)=\int^{v_{max}}_0 dv\gamma_{QCED} v^{\gamma_{QCED}-1}F_{YFS}(\gamma_{QCED})e^{\delta_{YFS}}\hat\sigma_{Born}((1-v)x_1x_2s)
\end{equation}
where we intend the well-known results for the 
respective parton-level Born cross
sections and the value of $v_{max}$ implied by the experimental cuts
under study. What is new here is the value for the QED$\otimes$QCD
exponent 
\begin{equation}
\gamma_{QCED}= \left\{2Q_f^2\frac{\alpha}{\pi}+2C_F\frac{\alpha_s}{\pi}\right\}L_{nls}
\label{expnt}
\end{equation}
where $L_{nls}=\ln x_1x_2s/\mu^2$ when $\mu$ is the factorization scale.
The functions $F_{YFS}(\gamma_{QCED})$ and $\delta_{YFS}(\gamma_{QCED})$
are well-known~\cite{yfs1} as well:
\begin{equation}
\begin{split}
F_{YFS}(\gamma_{QCED})&=\frac{e^{-\gamma_{QCED}\gamma_E}}{\Gamma(1+\gamma_{QCED})},\cr
\delta_{YFS}(\gamma_{QCED})&=\frac{1}{4}\gamma_{QCED}+(Q_f^2\frac{\alpha}{\pi}+C_F\frac{\alpha_s}{\pi})(2\zeta(2)-\frac{1}{2}),
\end{split}
\label{yfsfns}
\end{equation}
where $\zeta(2)$ is Riemann's zeta function of argument 2, i.e., $\pi^2/6$,
and $\gamma_E$ is Euler's constant, i.e., 0.5772... .
Using these formulas in (\ref{sigtot}) allows us to get the results
{\small
\begin{equation}
r_{exp}=
\begin{cases}
1.1901&, \text{QCED}\equiv \text{QCD+QED,~~LHC}\\
1.1872&, \text{QCD,~~LHC}\\
1.1911&, \text{QCED}\equiv \text{QCD+QED,~~Tevatron}\\
1.1879&, \text{QCD,~~Tevatron.}\\
\end{cases}
\label{res1}
\end{equation}}
We see that QED is at the level of .3\% at both LHC and FNAL.
This is stable under scale variations~\cite{CG1}.
We agree with the results in Refs.~\cite{baurall,ditt,russ,van1,van2}
on both of the respective sizes of the QED and QCD effects.
The QED effect is similar in size to structure function
results found in Refs.~\cite{cern2000,spies,james1,roth,james2},
for further reference.

We have shown that YFS theory (EEX and CEEX) extends to 
non-Abelian gauge theory and allows simultaneous exponentiation of 
QED and QCD, QED$\otimes$QCD exponentiation. For QED$\otimes$QCD we find that
full MC event generator realization is possible in a way that
combines our calculus with Herwig and Pythia in principle.
Semi-analytical results for QED (and QCD) threshold effects agree 
with literature on Z production. As QED is at the .3\% level, 
it is needed for 1\% LHC theory predictions. We have demonstrated
a firm basis for the complete ${\cal O}(\alpha_s^2,\alpha\alpha_s,\alpha^2)$
results needed for the FNAL/LHC/RHIC/ILC physics 
and all of the latter are in progress.
\section*{Acknowledgments}
One of us ( B.F.L.W.)
thanks Prof. W. Hollik for the support and kind
hospitality of the MPI, Munich, while a part of this work was
completed. We also thank Prof. S. Jadach for useful discussions.

\end{document}